\documentclass[11pt]{article}

\usepackage[latin1]{inputenc} 
\usepackage{latexsym}
\usepackage{graphicx}         
\usepackage{epsfig}           
\usepackage{amsmath}          
\usepackage{amsfonts}
\usepackage{fancyhdr}

\DeclareGraphicsExtensions{.eps,.ps,.pdf,.bmp}

\makeatletter

\begin{document}

\begin{center}
\lhead{{\bf\hspace*{1cm} 
\center{
	The International Conference on Statistical Methods \\
  and Econometric Models: Application in Economics and Finance \\
  \hspace{5.3cm} Tetouan, Morocco, February 24-25, 2020
} 
}}
\end{center}

\newtheorem{theorem}{Th\'eor\`em}[section]
\newtheorem{lemmma}{Lemma}[section]
\newtheorem{definition}[theorem]{D\'efinition}
\newtheorem{proposition}[theorem]{Proposition}
\newtheorem{corollary}[theorem]{Corollaire}
\newtheorem{remark}{\rm Remarque\/}
\newenvironment{prof}[1][Proof]{\noindent\textbf{#1.\ }}{\hfill~\rule{0.5em}{0.5em}}
\vspace{1cm}
\begin{center}
	{\fontsize{14}{30}\bf Econometric assessment of the monetary policy shocks in Morocco: Evidence from a Bayesian Factor-Augmented VAR
	}\end{center}
	
	\begin{center}
		\textbf{* Marouane Daoui}\\
	\end{center}
	
			\begin{center}
				Ph.D. in Economics and Management\\
				Faculty of Law, Economics and Social Sciences - Fez \\
				Sidi Mohamed Ben Abdellah University, Morocco \\  
				email:\{marouane.daoui@usmba.ac.ma\}
			\end{center}
		
\medskip
\vspace*{+0.2cm}

{\bf  Abstract:}
The analysis of the effects of monetary policy shocks using the common econometric models (such as VAR or SVAR) poses several empirical anomalies. However, it is known that in these econometric models the use of a large amount of information is accompanied by dimensionality problems. In this context, the approach in terms of FAVAR (Factor Augmented VAR) models tries to solve this problem. Moreover, the information contained in the factors is important for the correct identification of monetary policy shocks and it helps to correct the empirical anomalies usually encountered in empirical work. Following Bernanke, Boivin and Eliasz (2005) procedure, we will use the FAVAR model to analyze the impact of monetary policy shocks on the Moroccan economy. The model used allows us to obtain impulse response functions for all indicators in the macroeconomic dataset used (117 quarterly frequency series from 1985: Q1 to 2018: Q4) to have a more realistic and complete representation of the impact of monetary policy shocks in Morocco.

\vspace{2cm}

\section{Introduction}

In recent years, research in monetary economics has encouraged new thinking about how monetary policy can affect the economy, which has led to an evolution in the understanding of the monetary transmission mechanism. Similarly, macroeconomic and financial data are increasingly available, both in number and in length of series: : the conduct of monetary policy by central banks takes place in an information-rich environment. In terms of econometric tools, the methods used in the analysis of monetary policy shocks generally refer to standard VAR models (or their variants such as Vector Error Correction or Structural VAR).
\paragraph{}
The empirical studies demonstrate the strengths of VAR models in their classical form. Indeed, VAR models provide empirical responses of economic variables to a monetary policy shock that are consistent with the relative simplicity of the model, which is its main strength. However, the use of VAR models to analyze monetary policy shocks also has some limitations. A first limitation is the disagreement about the most appropriate strategy for identifying shocks. A second limitation arises from the small number of variables used to maintain a sufficient number of degrees of freedom for the estimation of the model. Similarly, the limited number of variables used in the VAR model is far from covering the hundreds of variables followed by economists and this fact can lead to two difficulties. %
First, the responses of variables to shocks may be biased due to the absence of a large number of variables containing information used in decision making. Second, an inescapable consequence of VAR models is that the number of observable responses is constrained by the small number of variables in the model. However, we could be interested in the responses of variables not included in the model (for reasons of degree of freedom). In response to the shortcomings of VAR models, Bernanke, Boivin and Eliasz (2005) propose new information-rich econometric models. These are Factor Augmented Autoregressive Vectors (FAVAR) based on the dynamic factor models (DFM) or diffusion indices used by Stock and Watson (2005).
\paragraph{}
The remainder of the document is organized as follows. The Literature Review section provides a comprehensive review of the relevant literature on the subject. The Data and Methodology sections provide a perspective on the data set used, as well as a brief introduction to the Bernanke, Boivin, and Eliasz (2005) approach used in this research. The results are presented and discussed with reference to the purpose of the study in the following section. The concluding section ends the paper with concluding remarks and policy implications.

\section{Literature review}
The monetary transmission takes place through multiple channels. Directly through the effect of interest rates on private consumption and investment decisions. Indirectly through the impact of exchange rates on import prices and external competitiveness, through the volume and price of credit granted by the financial system, through asset price adjustments and wealth effects and through private sector inflation expectations.
\paragraph{}
The propagation of monetary policy shocks through these channels depends on the development of financial markets, the consistency and credibility of the monetary policy regime, and the uncertainties of the domestic and external economic context.
\paragraph{}
We briefly summarize the results of some empirical studies, which use the VAR methodology and its variants, including FAVAR models, to assess the functioning of the interest rate channel of monetary policy. The reasoning behind this attempt is to achieve a better understanding of the macroeconomic implications of tightening cycles in monetary policy.
\paragraph{}
The interest rate transmission channel of monetary policy in advanced countries, as well as in emerging and developing countries, is a widely discussed topic. Mandler et al. (2016) use a Bayesian vector autoregression (BVAR) to model the dynamics of output, prices, and financial variables to an exogenous increase in the monetary policy interest rate of the European Central Bank (ECB) and the national central banks (NCBs) of the European Union member states (Eurosystem). The authors examine, in particular, the disparities in monetary transmission between the four major economies of the euro area (France, Germany, Italy and Spain). For this purpose, they show that real output reacts less negatively in Spain to monetary policy tightening than in the other three countries, while the reduction in the price level is smaller in Germany. At the same time, the results highlight that the rise in bond yields is stronger and more persistent in France and Germany than in Spain and Italy.
\paragraph{}
Georgiadis (2015), using the global vector autoregression (GVAR) model, reports the existence of asymmetries in the transmission of euro area monetary policy across different euro area countries using the global vector autoregression (GVAR) model. These asymmetries are mainly attributable to the structural characteristics of the economies (labor market rigidities, industrial structures, etc.). Eurozone economies with aggregate output linked to sectors where demand is sensitive to interest rates, real wages are higher and unemployment rigidities are low, have a stronger transmission of monetary policy to real activity.
\paragraph{}
Regarding emerging and developing countries, Cevik and Teksoz (2012), Jain-Chandra and Unsal (2014), and Anwar and Nguyen (2018) confirm the effectiveness of the interest rate channel in transmitting monetary policy shocks to the real economy using SVAR models.
\paragraph{}
The FAVAR literature on monetary transmission for countries with a fixed exchange rate regime is rather sparse. For example, Potjagailo (2016) implements a FAVAR model to explore the spillover effects of Eurozone monetary policy on thirteen EU countries outside the Eurozone. It is found that an expansionary Eurozone monetary policy shock increases output in most countries outside the Eurozone, while short-term interest rates and financial uncertainty decrease.
\paragraph{}
In their study, Daoui and Benyacoub (2021) employ the FAVAR model to investigate the effects of monetary policy shocks on economic growth in Morocco. Their analysis of the impulse response functions of some indicators of economic growth reveals that monetary policy shocks have a significantly negative impact on economic growth, as reflected by the overall decline in GDP.
\paragraph{}
In this context, our study contributes to the so far limited research on the FAVAR methodology that focuses on the interest rate channel of monetary policy in a small open economy like Morocco.

\section{Data}

Our data set consists of 117 quarterly macroeconomic time series for Morocco. The data sources include the statistics from Bank Al-Maghrib, High Commission for Planning, Foreign Exchange Office, Casablanca Stock Exchange, International Monetary Fund and World Bank. The data spans from first quarter of 1985 to last quarter of 2018. Due to the unavailability of some variables at quarterly frequency, the annual series have been quarterlyized, as have the monthly series. Quarterly values were interplotted using the sum (or mean) quadratic interpolation technique. In addition, the series that display seasonal variation were adequately adjusted. Furthermore, when plausible, we performed a logarithmic transformation to the series to achieve linearity. Also, non-stationary series have been transformed by taking the first difference (the second difference when plausible) and thus making them stationary. Decisions about these transformations are guided by unit root tests (ADF and Phillips Perron) combined with visual inspection and/or a priori economic judgment. Finally, since factor extraction could be affected by different units and scales, all series used to calculate the factors were standardized to have a mean of zero and a unit variance.
\paragraph{}
Following the procedure of Bernanke, Boivin, and Eliasz (2005), the data set is divided into fast-moving and slow-moving variables. Fast-moving variables respond contemporaneously to monetary policy, as they are highly sensitive to policy shocks. On the other hand, slow-moving variables are those that react with lags after a policy shock. Indeed, monetary aggregates, stock market indicators, interest rates, and exchange rates are considered "fast" variables. By contrast, the remaining variables are considered "slow" variables.

\section{Methodology}

In this section, the methodology presented is adapted from Bernanke, Boivin and Eliasz (2005). Our empirical application follows the so-called \textit{one-step} estimation (joint estimation) approach to estimating monetary FAVARs.
\paragraph*{}
As discussed earlier, in FAVAR models, the information in a large data set is summarized by a few variables called factors, which are incorporated into a VAR model. This allows us to expand the dataset used in standard VAR models and generate the response of hundreds of variables to monetary policy shocks. The joint dynamics of $(F_t,\ Y_t)$ are assumed by Bernanke, Boivin and Eliasz (2005) to be given by the following equation:
\begin{equation}
	\left[ {\begin{array}{*{20}{c}}
			{{F_t}}\\
			{{Y_t}}
	\end{array}} \right] = {\rm{\Gamma }}\left( L \right)\left[ {\begin{array}{*{20}{c}}
			{{F_{t - 1}}}\\
			{{Y_{t - 1}}}
	\end{array}} \right] + {\nu _t}\;\;\;\;:\;\;\forall \;t = 1, \ldots ,T \label{eq1}
\end{equation}

where $\Gamma(L)$ is a conformable lag polynomial of finite order $d$ and the error term, ${\nu _t}$, is mean zero with covariance matrix $\Sigma$. The vector ${Y_t}$ contains $M$ observable economic variables and the vector ${F_t}$ represents $K$ unobserved factors that are assumed to influence the economic variables. Factors can be considered as unobservable concepts (such as economic activity or investment climate) that cannot be represented by any macroeconomic observable series, but by several series of economic indicators. Subsequently, if the terms in $\Gamma(L)$ that relate ${Y_t}$ to ${F_{t - 1}}$ are all zero, equation (1) would reduce to a standard VAR in ${Y_t}$. If ${Y_t}$ is related to the lagged factors, then equation (1) would be called a Factor Augmented Vector Autoregression (FAVAR).
\paragraph{}
Since the ${F_t}$ factors are unobservable, equation (1) cannot be estimated directly. However, if we interpret the factors as representing forces that potentially affect many economic variables, we can hope to infer something about the factors from observations on a variety of economic time series. To make this concrete, suppose we have a number of informational time series, denoted by the $N \times 1$ vector $X_t$.
\paragraph{}
Therefore, the number of time series $N$ in $X_t$ is assumed to be large, and may well be larger  than the number of periods $T$. Bernanke, Boivin, and Eliasz (2005) suppose that the time series in $X_t$ is related to unobservable factors Ft and observable economic variables $Y_t$ through the following equation:
\begin{equation}
	{X_t} = {\lambda _f}{F_t} + {\lambda_y}{Y_t} + {e_t}\;\;\;\;:\;\;\forall \;t = 1, \ldots ,T \label{eq2}
\end{equation}
where ${\lambda _f}$ is $N \times K$ matrix of factor loadings, ${\lambda_y}$ is $N \times M$ and ${e_t}$ is a $N \times 1$ vector of error terms that are assumed to be zero-mean but may exhibit a small level of cross-correlation depending on the estimation method (estimation is by principal components or likelihood methods). The idea that $Y_t$ and $F_t$, which may be correlated, summarize the common forces driving the dynamics of the noisy measurements of $X_t$ is expressed by equation (2). 
\paragraph{}
In addition, in order to jointly estimate equations (1) and (2) by likelihood methods, a transformation of the model into a state-space form is required. Moreover, the factors are effectively identified, in this method, by both the observation equation and the transition equation of the state-space model. Here, the identification instrument is to make an assumption that constrains the channels through which $Y$ affect $X$ contemporaneously. For this purpose, joint likelihood estimation requires only that the first $K$ variables in the data set be chosen from the set of "slow moving" variables and that the recursive structure be imposed in the transition equation. The identification scheme employed is developed by Bernanke, Boivin and Eliasz (2005) in detail.
\paragraph{}
Moreover, we assume that the policy rate is the only observable factor, that is the only variable included in $Y_t$. In doing so, we treat the policy rate as a factor and interpret it as the instrument of monetary policy. This is based on the assumption that monetary policy has a pervasive effect on the economy.
\paragraph{}
The likelihood-based method is fully parametric and computationally more demanding. While in principle the estimation of equations (1) and (2) jointly by Maximum Likelihood is possible, assuming independent normal errors, Bernanke, Boivin and Eliasz (2005) argue that it is infeasible in practice due to the very large size of this model and the irregular nature of the likelihood function.
\paragraph{}
Bernanke, Boivin, and Eliasz (2005) implement a multi-move version of the Gibbs sampler in which the factors are sampled conditionally on the most recent factor draws. This Bayesian approach is attempted to overcome the high dimensionality problem of the model by approximating marginal likelihoods with empirical densities.

\section{Results and discussion}

\begin{figure}[!h]
	\begin{center}
		\includegraphics[width=0.7\textwidth]{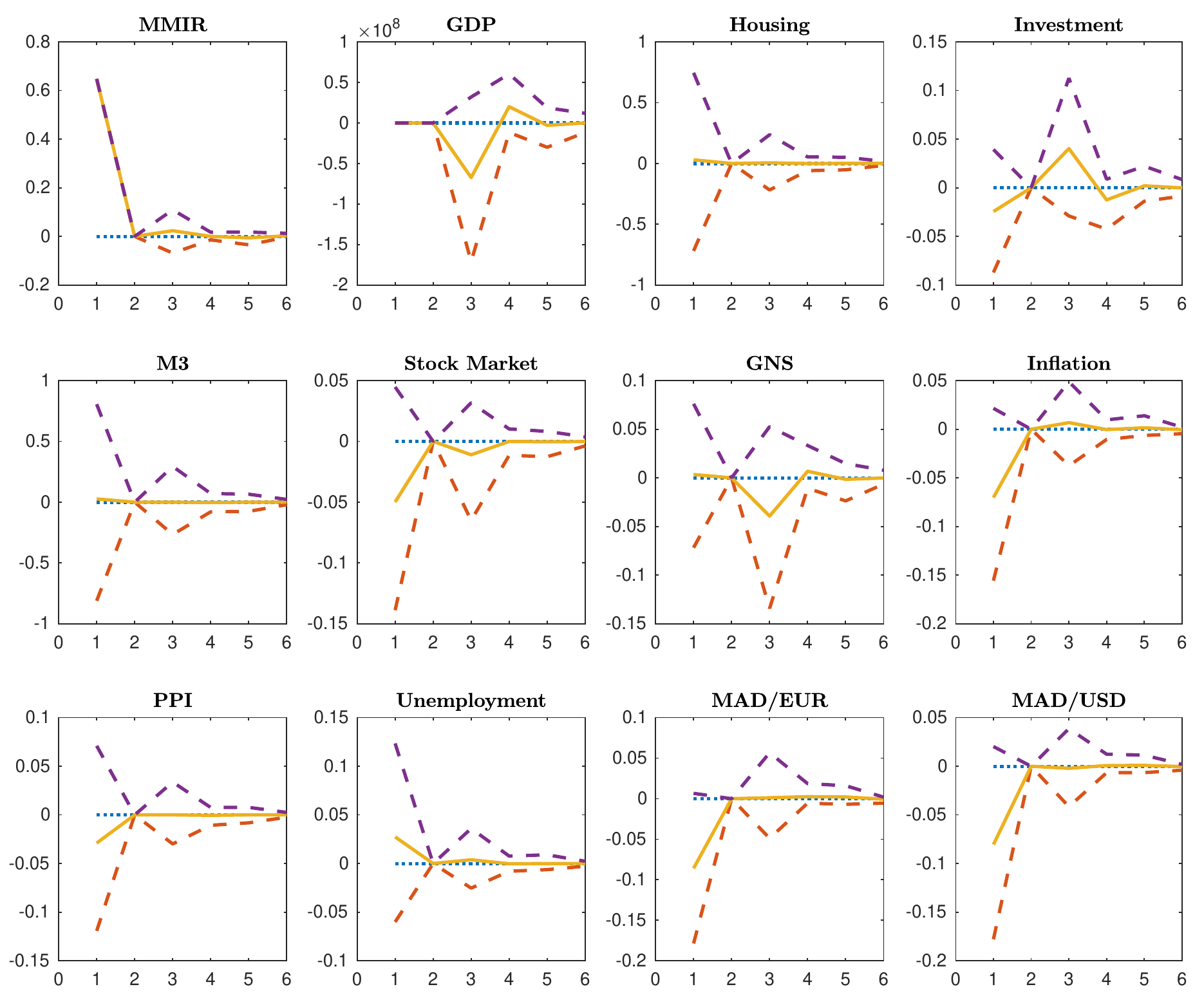}\\
{\footnotesize \textbf{Source :} Author's simulations.}
\caption{Impulse response functions of a set of variables related to the Moroccan economy to a contractionary monetary policy shock (from a 0.25 p.p. increase in the money market interest rate).} \label{fig_rep_mmir}
	\end{center}
\end{figure}

\begin{figure}[!h]
	\begin{center}
	\includegraphics[width=0.7\textwidth]{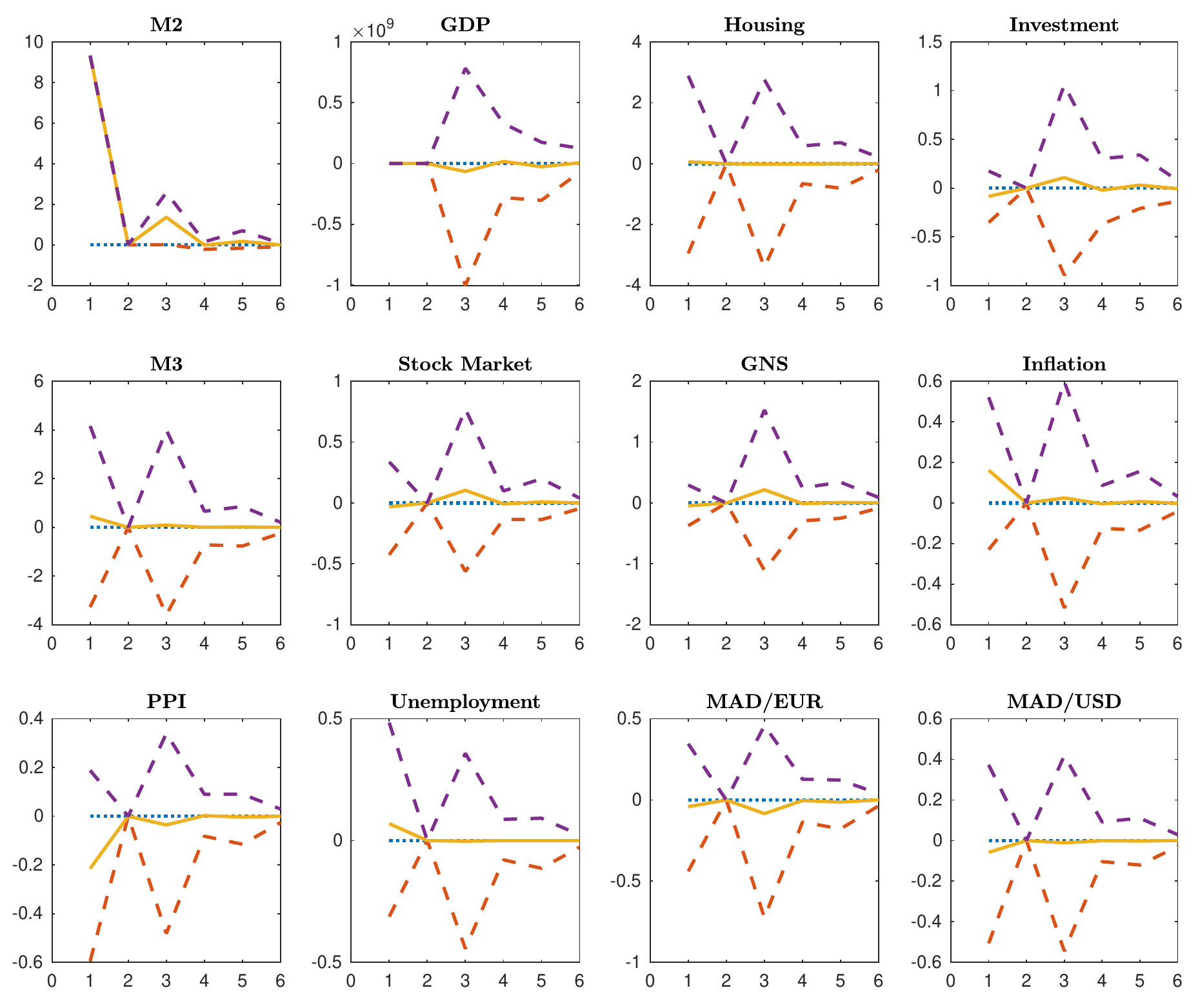}\\
{\footnotesize \textbf{Source :} Author's simulations.}
	\caption{Impulse response functions of a set of variables related to the Moroccan economy to a contractionary monetary policy shock (from a 0.25 p.p. increase in the M2 money supply).} \label{fig_rep_m2}
\end{center}
\end{figure}

The FAVAR model used is designed to examine the magnitude of the effects of monetary policy shocks. The responses of selected macroeconomic variables indicative of economic activity to a 0.25 percentage point increase in the money market interest rate (MMIR) and the M2 money supply are presented in Figures (\ref{fig_rep_mmir}) and (\ref{fig_rep_m2}), respectively. The dashed lines represent the 95\% confidence intervals around the impulse response functions.
\paragraph{}

For an MMIR shock, our results suggest that a tightening of monetary policy leads to a small decline in gross domestic product (GDP), producer price index (PPI) and the exchange rate (MAD/EUR). In contrast, we observe an increase in investment, stock market capitalization and gross national savings (GNS). However, the remaining variables do not represent any significant response.
\paragraph{}
For a money supply shock M2, our results show an increase in investment, stock market capitalization, gross national savings and producer price index. However, a decrease in M3, inflation and unemployment is observed. The other variables show a very slight response.
\paragraph{}
Insignificant impulse responses of a few variables could be a particular problem with the sample, especially keeping in mind the structural changes in the study period and the underlying instability of the factor loadings, which makes it difficult to reach more robust conclusions.

\paragraph{}

Having an idea about the number of factors needed to capture the information essential for adequate modeling of the effects of monetary policy is an important practical issue to consider when using the FAVAR methodology. Therefore, we investigated the sensitivity of the impulse response functions to an alternative number of factors. Thus, we estimated the corresponding FAVAR model with a different number of factors to demonstrate how the inclusion of factors can improve the results. As such, we found that the results were not robust to the use of more than three factors. We also experimented with the number of lags included in the model. In this regard, FAVAR with 4 lags produces the most economically and statistically reasonable impulse responses.
\paragraph{}
Overall, our baseline specification (FAVAR model with 3 factors and 4 lags) appears to provide a consistent and sensitive measure of the effect of monetary policy.
\paragraph{}
Finally, while most empirical studies support that the FAVAR approach results in better empirical estimates, we still cannot rule out the possibility of a potential mis-measurement due to the exclusion of some series. Moreover, in the current highly globalized world, central banks draw on both the economic state of the country and what is happening elsewhere. Such variables as commodity price indices as well as foreign interest rates can have a potential effect on the transmission of monetary policy. It may be especially interesting to introduce foreign interest rates as exogenous monetary policy shocks. This could be the subject of future studies.

\section{Conclusion}

This study examines the effectiveness of the interest rate policy transmission mechanism in Morocco using the Bayesian Factor-Augmented Vector Autoregressive (FAVAR) approach pioneered by Bernanke, Boivin and Eliasz (2005). Moreover, this paper tries to fill a gap in the literature for Morocco as it assesses the effects of monetary policy in a data-rich environment. At the same time, this study examines the challenges faced by monetary authorities in small open economies in understanding how monetary policy instruments affect the economy.
\paragraph{}
Our empirical application of the FAVAR methodology shows that monetary tightening induces a decline in in gross domestic product, producer price index and the exchange rate MAD/EUR. In contrast, the following increase: investment, stock market capitalization and gross national savings.
\paragraph{}
However, some impulse responses are statistically insignificant, most likely as a result of a particular problem with the sample, especially when taking into account the structural changes that occurred over the period of analysis and the associated instability of factor loadings, which does not allow for more robust conclusions. Therefore, these results should be considered indicative, rather than complete and absolute. In addition, the possibility of a possible mis-measurement due to the exclusion of some series may also arise.
\paragraph{}
The results corroborate the empirical literature and economic theory, thus providing a more complete view of the transmission mechanism and the effect of monetary policy on the Moroccan economy, which could therefore be useful and interesting for monetary authorities when designing and conducting monetary policy.

\bibliographystyle{plain}


\clearpage
\newpage
\appendix

\section*{Appendix: Data Description}
{\footnotesize
The database is composed of 117 time series at quarterly frequency from 1985:Q1 to 2018:Q4. 'Fast' variables are denoted with an asterisk (*), the remaining block of variables is considered 'slow'. The transformation of the variables to make them stationary is done according to the transformation codes (TC) below:\\
(1) No transformation: $X_{it}=Y_{it}$\\
(2) First difference: $X_{it}=\Delta Y_{it}$\\
(3) Second difference: $X_{i\mathrm{t}}=\Delta^{2}Y_{i\mathrm{t}}$\\
(4) Logarithm $X_{i\mathrm{t}}=\log Y_{i\mathrm{t}}$\\
(5) First difference of logarithm: $X_{i\mathrm{t}}=\Delta\log Y_{i\mathrm{t}}$\\
(6) Second difference of logarithm: $X_{it}=\Delta^{2}\log Y_{it}$
}

\includegraphics[scale=1]{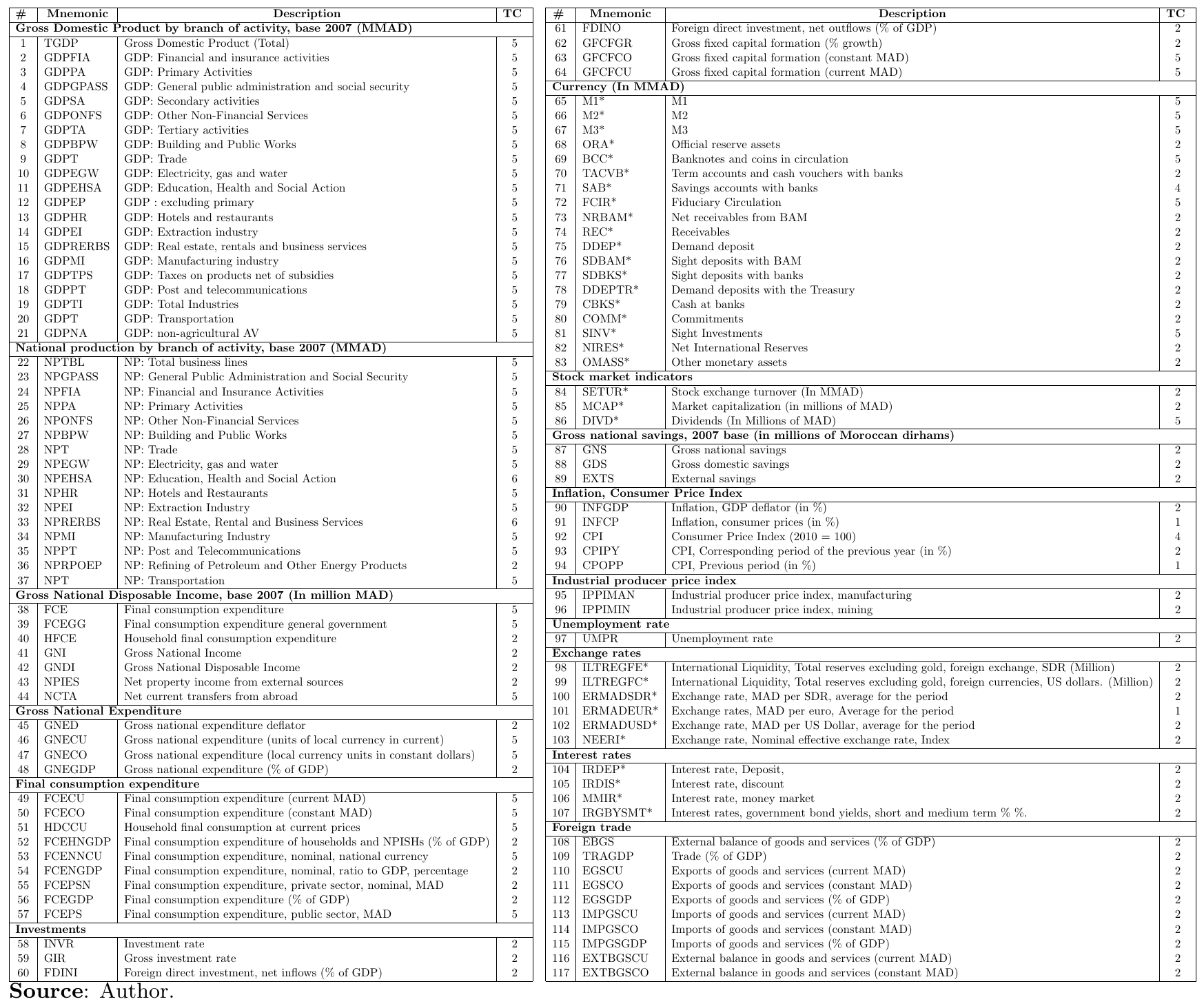}

\end{document}